\documentclass[final]{svjour2}
\usepackage{graphicx}
\usepackage{amssymb}
\usepackage{mathptmx}
\usepackage[numbers]{natbib}
\usepackage{dcolumn}
\usepackage{color}
\usepackage{bm}
\begin{document}

%%%%%%%%%%%%%%%%%%%%%%%%%%%%%%%%%%%%%% AUTHORS %%%%%%%%%%%%%%%%%%%%%%%%%
\author{E.V. Kozik \and B.V. Svistunov}
\institute{E.V. Kozik \at Theoretische Physik, ETH Z\"urich, 8093 Z\"urich, Switzerland
\and
B.V. Svistunov \at Department of Physics, University of Massachusetts, Amherst, MA 01003, USA \\
Russian Research Center ``Kurchatov Institute'', 123182 Moscow, Russia}

%%%%%%%%%%%%%%%%%%%%%%%%%%%%%%%%%%%%%%%%%%%%%%%%%%%%%%%%%%%%%%%%%%%%%%%%%%%%%%

\title{Comment on ``Symmetries and Interaction Coefficients of Kelvin waves'' by Lebedev and L'vov}

%%%%%%%%%%%%%%%%%%%%%%%%%%%%%%%%%%%%%%%%%%%%%%%%%%%%%%%%%%%%%%%%%%%%%%%%%%%%%%
\date{XX.XX.2010}
\maketitle

\begin{abstract}
We comment on the claim by Lebedev and L'vov \cite{Lebedev_L'vov} that the symmetry with respect to a tilt of a quantized vortex line does not yet prohibit coupling between Kelvin waves and the large-scale slope of the line. Ironically, the counterexample of an effective scattering vertex in the local induction approximation (LIA) attempted by Lebedev and L'vov invalidates their logic all by itself being a notoriously known example of how symmetries impose stringent constraints on kelvon kinetics---not only the coupling in question but the kinetics in general are absent within LIA.  We further explain that the mistake arises from confusing symmetry properties of a specific mathematical representation in terms of the canonical vortex position field $w(z)=x(z)+iy(z)$, which explicitly breaks the tilt symmetry due to an arbitrary choice of the $z$-axis, with those of the real physical system recovered in final expressions.
\end{abstract}

\vspace{1cm}

In their paper \cite{Lebedev_L'vov} Lebedev and L'vov suggest what they believe to be a counterexample for the symmetry argument \cite{talk, KS_sym}, which leads to the proof of locality of the Kelvin-wave cascade in the wavenumber space. The cascade locality, which is a crucial ingredient of the Kelvin-wave kinetic theory \cite{KS_04, KS_JLTP}, was claimed to be violated in the recent paper by Laurie, \textit{et al.} \cite{LN_nonlocal} on the grounds of a direct, and quite cumbersome, analytical evaluation of the collision term. In particular it was argued that the leading contribution to kinetics at a given wavenumber scale $k$ is due to coupling to the largest scales $R_0$, $k \gg R_0^{-1}$, whereas the long-wavelength modes enter the collision term via the square average \textit{angle} of vortex-line distortion at the scale $R_0$. However, the present authors pointed out a general symmetry argument (presented in detail in Ref.~\cite{KS_sym}) invalidating the calculation of Laurie, \textit{et al.} and immediately proving the locality of the Kelvin-wave cascade. The argument is based on the geometrical nature of the canonical complex field $w(z)=x(z)+iy(z)$ describing the distortions of a vortex line in the Cartesian coordinate system: due to the symmetry of the kelvon Hamiltonian \cite{Sv95} with respect to the global tilt of the vortex line, the short-wave kinetics can \textit{not} couple to the large-scale angle $w'$, but only to higher derivatives of the field $w$ at the scale $R_0$ with the main contribution coming from the large-scale curvature $w''$.  

The example suggested by Lebedev and L'vov exposes the essence of their misconception of the role of symmetries in the problem. They consider the expression for the total line length in terms of the variable $w(z)$, $L=\int \sqrt{1+|w'(z)|^2} dz$, which is trivially invariant under the tilt of the vortex line and can be thought of as some effective Hamiltonian. However, they observe that the explicit form of the kelvon scattering vertices $T^{3,4}_{1,2} \propto k_1 k_2 k_3 k_4$, $W^{4,5,6}_{1,2,3} \propto k_1 k_2 k_3 k_4 k_5 k_6$, obtained by the Taylor expansion of $L$ with respect to $|w'(z)| \ll 1$ and a Fourier transform of the field $w(z)$, $w_k$, $k=k_1, \ldots, k_6$, is inconsistent with the ``naive'' symmetry argument since these vertices do have a linear dependence on all the wavenumbers, which leads to an explicit dependence on the large-scale slope despite the tilt symmetry if, say, $k_1 \ll k_{2,\ldots,6}$. In fact, the naiveness of the authors of Ref.~\cite{Lebedev_L'vov} is in failing to appreciate that, being intermediate auxiliary concepts, the bare vertices themselves have no direct physical meaning and thus 
are not supposed to respect the tilt invariance on the individual basis. It is only when $T^{3,4}_{1,2}$ and $W^{4,5,6}_{1,2,3}$ are combined into the full physical six-wave scattering vertex $V^{4,5,6}_{1,2,3}$---in accordance with Ref.~\cite{KS_04} the four-wave vertex $T^{3,4}_{1,2}$ contributes to $V^{4,5,6}_{1,2,3}$ along with $W^{4,5,6}_{1,2,3}$ in the second order with a virtual kelvon emission/absorption process---that one can speak of restrictions imposed on kelvon kinetics by the symmetries via $V^{4,5,6}_{1,2,3}$. 

Curiously, the example picked by Lebedev and L'vov is a perfect illustration of how symmetries determine the kinetics specifically. Despite the linear in $k_{1, \ldots, 6}$ dependence of the vertices $T^{3,4}_{1,2}$ and $W^{4,5,6}_{1,2,3}$, not only $V^{4,5,6}_{1,2,3}$ is independent of the local large-scale slope, but it is \textit{exactly equal to zero} \cite{KS_04}, because $L$ in terms of $w(z)$ is nothing but the Hamiltonian of the local induction approximation (LIA) \cite{Sv95}, which is known to feature a considerably wider class of symmetries then the full Biot-Savart model. More specifically, LIA has an \textit{infinite} number of constants of motion, which immediately require  $V^{4,5,6}_{1,2,3} \equiv 0$. This fact is also revealed by a direct calculation of $V^{4,5,6}_{1,2,3}$ \cite{KS_04, KS_JLTP}. Analogously, in a more general case of the full Biot-Savart model, the invariance of dynamics with respect to the shift and tilt of the vortex line prescribes that, after combining the corresponding vertices $\tilde{T}^{3,4}_{1,2}$ and $\tilde{W}^{4,5,6}_{1,2,3}$ into $\tilde{V}^{4,5,6}_{1,2,3}$, the terms $\propto k_1 k_2 k_3 k_4 k_5 k_6$ at $k_1 \ll k_{2,\ldots,6}$ must cancel exactly leaving the first non-vanishing contribution to $\tilde{V}^{4,5,6}_{1,2,3}$ proportional to $k_1^2$, which was demonstrated in our Ref.~\cite{KS_sym}. This was not the case in the calculation by Laurie, \textit{et al.} \cite{LN_nonlocal} and in the following non-local Kelvin-wave cascade scenario by L'vov and Nazarenko \cite{LN}; a discussion of the possible source of this mistake can be found in Ref.~\cite{KS_sym}. 

Finally, we note that in Ref.~\cite{KS_sym} we explicitly derive an effective Lagrangian of the short-wave modes superimposed on a vortex with a large-scale curvature. The advantage of our approach over similar calculations presented in Ref.~\cite{Lebedev_L'vov} is in parametrising the fast short-wave field by the arc length of the long-wave distortion rather then by the coordinate $z$ since the latter requires to account for corrections to the vortex-line length from the large-scale distortion as well. Our derivation explicitly demonstrates that the coupling to the long-wave field is exclusively via the large-scale curvature thereby leading to the proof of locality of the Kelvin-wave cascade.

\end{document}